\documentclass[journal]{IEEEtran}

\usepackage{cite}
\usepackage{dblfloatfix}
\usepackage{mathtools}
\usepackage{float}
\usepackage{amsmath,amssymb,amsfonts}
\usepackage{algorithm}
\usepackage[noend]{algpseudocode}
\usepackage{graphicx}
\graphicspath{{./Figures/}}
\usepackage{textcomp}
\usepackage{comment}
\usepackage{gensymb} 
\usepackage{booktabs}
\usepackage{adjustbox}
\usepackage{siunitx}
\usepackage{bm} 
\usepackage{cite} 
\usepackage{cleveref} 
\usepackage{url}
\usepackage{array,multirow}
\usepackage[table]{xcolor}
\usepackage{multirow,tabularx} 
\usepackage{xcolor,colortbl}
\usepackage[skins]{tcolorbox}
\def\BibTeX{{\rm B\kern-.05em{\sc i\kern-.025em b}\kern-.08em
T\kern-.1667em\lower.7ex\hbox{E}\kern-.125emX}}
\newcolumntype{L}[1]{>{\raggedright\let\newline\\\arraybackslash\hspace{0pt}}m{#1}}
\newcolumntype{C}[1]{>{\centering\let\newline\\\arraybackslash\hspace{0pt}}m{#1}}
\newcolumntype{R}[1]{>{\raggedleft\let\newline\\\arraybackslash\hspace{0pt}}m{#1}}
\usepackage{enumitem}

\begin{document}

%


\title{Under-frequency Load Shedding for Power Reserve Management in Islanded Microgrids}

\author{Bei~Xu,~\IEEEmembership{Student Member,~IEEE,}
        Victor~Paduani,~\IEEEmembership{Member,~IEEE,}   
        Qi~Xiao,~\IEEEmembership{Student Member,~IEEE,}   
        Lidong Song,~\IEEEmembership{Member,~IEEE,} 
        David~Lubkeman,~\IEEEmembership{Fellow,~IEEE,}   
        and~Ning~Lu,~\IEEEmembership{Fellow,~IEEE}
\thanks{This research is supported by the U.S. Department of Energy's Office of Energy Efficiency and Renewable Energy (EERE) under the Solar Energy Technologies Office Award Number DE-EE0008770. The authors are with the Department of Electrical and Computer Engineering, North Carolina State University, Raleigh, NC, 27606 USA. E-mails: (bxu8@ncsu.edu and nlu2@ncsu.edu).
}
}

\maketitle

\begin{abstract}
This paper introduces under-frequency load shedding (UFLS) schemes specially designed to fulfill the power reserve requirements in islanded microgrids (MGs), where only one grid-forming resource is available for frequency regulation.
When the power consumption of the MG exceeds a pre-defined threshold, the MG frequency will be lowered to various setpoints, thereby triggering UFLS  for different levels of load reduction. Three types of controllable devices are considered for executing UFLS: sectionalizers, smart meters, and controllable appliances. 
To avoid unnecessary UFLS activation, various time delay settings are analyzed, allowing short-lived power spikes caused by events like motor startups or cold-load pickups to be disregarded.
We tested the proposed UFLS schemes on a modified IEEE 123-bus system on the OPAL-RT eMEGASIM platform. Simulation results verify the efficacy of the proposed approaches in restoring power reserves, maintaining phase power balance, and effectively handling short-lived power fluctuations. 
Furthermore, in comparison to sectionalizer-based UFLS, using smart meters or controllable loads for UFLS allows for a more accurate per-phase load shedding in a progressive manner. As a result, it leads to better balanced three-phase voltage and serves more loads.

\end{abstract}

\IEEEpeerreviewmaketitle

\section{Introduction}

\IEEEPARstart{W}{hen} operating islanded microgrids (MGs), grid-forming (GFM) sources are required to establish and maintain voltage and frequency while meeting power reserve requirement (PRR) at all times to effectively manage sudden load increases. These GFM functions play a pivotal role in ensuring reliable and stable MG operation\cite{som2023bess,karimi2016new}.

In a small MG, the absence of load diversity leads to frequent large load fluctuations, making the fulfillment of PRR vital for ensuring a reliable and stable MG operation~\cite{li2019temporally}.
Nevertheless, depending solely on GFM resources for maintaining power reserves in an islanded MG poses various challenges, particularly when there is only one GFM resource available. First, variable renewable generation resources, such as photovoltaic (PV) and wind, often exhibit considerable fluctuations in their power outputs. As a result, the GFM resource must rapidly adapt its output to accommodate these unpredictable power variations and ensure an adequate power reserve margin. However, this frequent adjustment can lead to increased wear-and-tear\cite{ke2014control} and a degradation of the GFM's capabilities to regulate frequency and voltage. Second, MG reconfiguration can cause significant power surges due to cold-load pickup and motor startup events, further complicating the GFM capability for meeting PRR while regulating large voltage and frequency changes. Thus, to ensure a more stable and resilient MG operation, it is essential to explore alternative approaches and complementary resources for meeting PRR, rather than relying solely on a single GFM resource. 

By integrating advanced forecasting algorithms and utilizing battery energy storage systems (BESS), grid-following (GFL) wind and solar farms can effectively contribute to power reserves\cite{chang2021coordinated,li2013battery,nair2021analysis}. Thus, in MGs with multiple GFM resources, an option exists to alleviate overloading by switching the overloaded resources from GFM to GFL mode\cite{rezkallah2020implementation,bayhan2018simple}. Another promising approach is to aggregate distributed energy resources (DERs) into a virtual power plant (VPP), allowing the VPP to serve as a source of power reserves\cite{naughton2021co,lee2021optimal}. However, note implementing the aforementioned methods requires the presence of multiple GFM or GFL resources within an islanded MG. Therefore, in a MG with just a single GFM resource, and where distributed generation resources like rooftop PV systems lack GFM capabilities, demand response (DR) emerges as the sole viable solution.  

Over the last decade, significant advancements in communication and control technologies have rendered DR a viable and effective resource for providing grid regulation and load following services \cite{lu2012design,wu2017hierarchical,sun2017modeling}. By promptly responding to price or load balancing signals, controllable loads can be efficiently activated and deactivated, allowing for real-time adjustments to electricity consumption
\cite{biegel2016sustainable,junhui2021optimal}. However, fulfilling PRR demands exceptionally rapid responses from DR resources. Achieving this level of responsiveness necessitates the real-time transmission of price or control signals to hundreds or even thousands of controllable loads, which heavily relies on highly reliable communication networks. Unfortunately, this can become impractical during prolonged outages.

\begin{table*}[htb]
    \centering
    \vspace{-10pt}
    \caption{Comparison of Existing and Proposed Under-Frequency Load Shedding Schemes}
    \vspace{-5pt}
    \begin{adjustbox}{max width=.98\textwidth}
    \begin{tabular}{C{1.5cm}|C{1.3cm}|C{1.5cm}|C{2.0cm}|C{2.5cm}|C{3.0cm}|C{1.2cm}|C{1.2cm}}
        \hline
        \hline
        Method & Object & Operation Condition & Triggered by & UFLS Execution & Control Mechanism & 3-Phase Imbalance & Power Surge \\
        \hline
        \hline
        \vspace{0.15cm} \multirow{2}{1.5cm}{\centering Traditional UFLS} &\multirow{2}{1.3cm}{\centering Recover system frequency} & \vspace{0.05cm} \multirow{2}{1.5cm}{\centering Emergency response} & \multirow{2}{2.0cm}{\centering Large frequency drops due to outages}  & UFLS relays\cite{potel2017under,gordon2021impact,banijamali2018semi} & Autonomous\cite{potel2017under,gordon2021impact,banijamali2018semi} &  \vspace{0.15cm} \multirow{2}{1.2cm}{\centering No} &  \vspace{0.15cm} \multirow{2}{1.2cm}{\centering No} \\
        \cline{5-6}
         & & & & Controllable loads\cite{zhou2018two,li2019continuous,gu2014adaptive,karimi2016new} & Centralized\cite{zhou2018two,li2019continuous,karimi2016new} \newline Decentralized\cite{gu2014adaptive} & & \\
        \hline
        \vspace{0.25cm} \multirow{2}{1.5cm}{\centering \textbf{Proposed Method}} & \multirow{2}{1.3cm}{\centering Keep power reserve margin} & \vspace{0.3cm} \multirow{2}{1.5cm}{\centering Normal operation} & \vspace{0.3cm} \multirow{2}{2.0cm}{\centering Low power reserve} & \centering Sectionalizers & \vspace{0.3cm} \multirow{2}{3.0cm}{\centering Autonomous} & No & \vspace{0.3cm} \multirow{2}{1.2cm}{\centering Yes} \\
        \cline{5-5} \cline{7-7}
         & & & & Smart meters; Controllable appliances & & Yes & \\
        \hline
    \end{tabular}    
    \end{adjustbox}
    \vspace{-10pt}
    \label{tab:SUM1} 
\end{table*}

Therefore, in this paper, we introduce a pioneering method for delivering power reserve through under-frequency load shedding (UFLS). The key innovation of this approach lies in using a power reserve threshold as the triggering mechanism while employing modulated system frequency as control signals to enable autonomous DR for meeting PRR. The primary advantage of this approach is its capability of operating an islanded MG during prolonged outages without relying on reliable communication networks. Using frequency as the control signal, the method is immune to communication failures, which greatly increases the MG's robustness. 
The second contribution of the paper is the introduction of an intricate autonomous UFLS mechanism for achieving three critical objectives: 1) uniformly distributed device tripping delays for progressively shedding loads when meeting the Power-to-Remain-Ratio requirement to minimize the amount of unserved loads and ensure efficient response, 2) a ramp rate checking for avoiding inadvertent load shedding in response to short-lived power surges to prevent unnecessary load interruptions, and 3) uniformly distributed device recovery delays to avoid large power fluctuations caused by simultaneously turning on DR resources.
By implementing this sophisticated delay mechanism, our approach further enhances the reliability and precision of UFLS, making it a robust solution for power reserve management in islanded MGs.

A comprehensive comparison between conventional UFLS and the proposed UFLS can be found in Table \ref{tab:SUM1}. Traditionally, UFLS is used as an emergency response mechanism to prevent frequency collapse in large-scale power systems\cite{wen2017enhancing}. In the main grid, system frequency declines when demand surpasses generation during outages\cite{ferraro2018stochastic}. When the system frequency drops below a predetermined level, UFLS will be triggered to quickly disconnect a large amount of loads to regain balance between the demand and supply\cite{gordon2021impact,zhou2018two}. 

In contrast, the proposed UFLS method prioritizes the preservation of power reserves in an islanded MG during regular operation to meet the PRR. Consequently, UFLS activation is based on a power threshold determined by the PRR, rather than relying on frequency drops caused by insufficient generation. Furthermore, we modulate the system frequency as universal control signals to  activate distributed controllable devices within the MG, such as sectionalizers, smart meters, or appliances, for autonomous power reserve management. 
To guarantee the stable and efficient management of power reserves, the UFLS mechanism incorporates ramp rate checking and uniformly distributed tripping and recovery delays, which can prevent unnecessary triggering during motor start-ups, and synchronized activation of DR devices.
Moreover, employing per-phase UFLS enables the effective correction of three-phase imbalances, which is a novel application of UFLS further showcasing the versatility and potential of UFLS in addressing MG power management challenges.

The subsequent sections of the paper are arranged as follows: Section II introduces the methodology, Section III presents the simulation results, and Section IV concludes the paper while suggesting directions for future research.

\section{Methodology}\label{Section2}
In this section, we first present a simplification made to the conventional BESS GFM control structure to facilitate modulating system frequency for providing a MG-wise UFLS control signal that enables autonomous UFLS response. Then, we introduce three UFLS schemes: sectionalizer based, smart meters based, and appliance-based.

\vspace{-5pt}
\subsection{Modification to the BESS-based GFM Control}
\begin{figure}[htb]	
    \vspace{-10pt}
    \centerline{\includegraphics[width=0.47\textwidth]{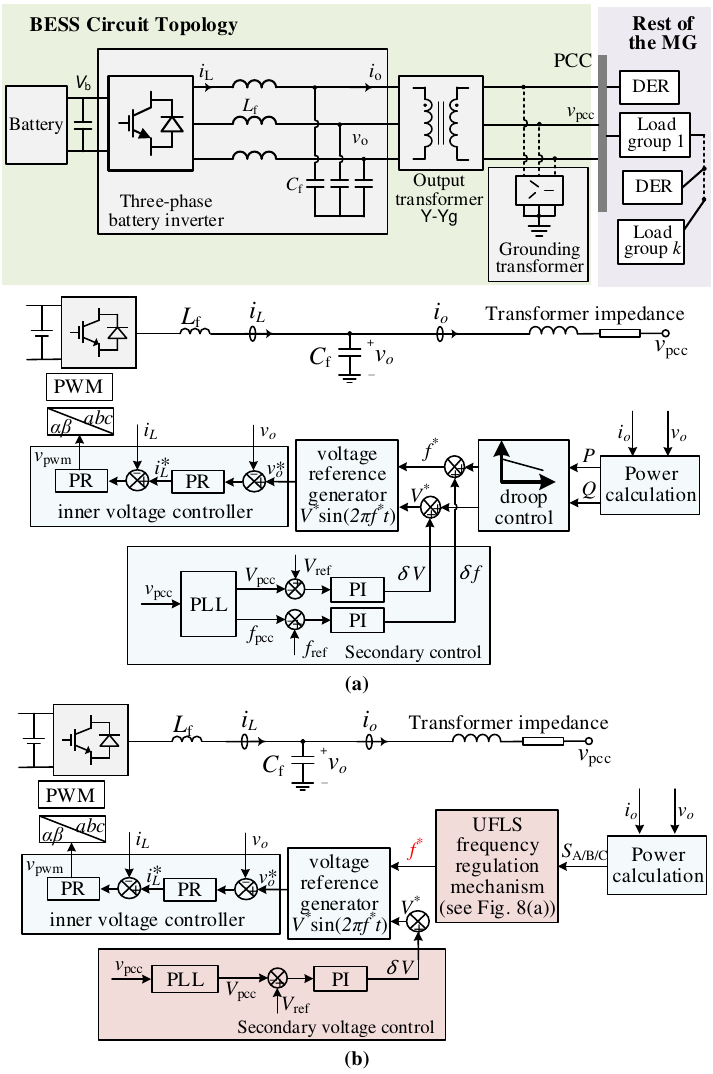}}
    \vspace{-12pt}
    \caption{(a) BESS circuit topology and conventional BESS GFM control structure, and (b) The proposed, simplified BESS control structure. }
    \label{fig:control}
\end{figure}

The MG test system utilized in this study is depicted in Fig. \ref{fig:control}(a). The sole GFM resource is a BESS, comprising a three-phase inverter, an LC filter, and a Y-Yg isolation transformer that connects the BESS to the main grid. A grounding transformer is added to mitigate the zero-sequence components for keeping the voltage at the point of common coupling (PCC), $v_\mathrm{pcc}$, balanced when serving highly unbalanced single-phase loads \cite{xu2022novel}. 
It is essential to highlight that the proposed UFLS mechanism is employed to fulfill PRR in an islanded MG with just one GFM resource.

The conventional BESS-based GFM control often applies the droop-based hierarchical approach, originally introduced by Guerrero \emph{et al.}\cite{standard_2010hierarchical}, as illustrated in Fig. \ref{fig:control}(a). This control structure consists of inner current and voltage control loops, primary (droop) control, and secondary control. When operating in off-grid mode, the GFM BESS inverter functions as a voltage source with two main objectives: regulating the PCC voltage ($v_\mathrm{pcc}$) and establishing the system frequency ($f$).

The inner current and voltage controller play a vital role in ensuring the regulation of frequency and inverter output voltage ($v_\mathrm{o}$). The droop control is responsible for adjusting the frequency reference ($f^*$) and the voltage reference amplitude ($V^*$) based on the active and reactive powers ($P$ and $Q$) using the well-known P/Q droop method. Meanwhile, the secondary control is aimed at restoring the frequency and voltage. This control algorithm enables power sharing regulation among multiple GFM resources and facilitates a seamless transition between off-grid mode and grid-connected mode without the need to modify the control structure.

\begin{figure}[t]  
\vspace{-10pt}
    \centerline{\includegraphics[width=0.44\textwidth]{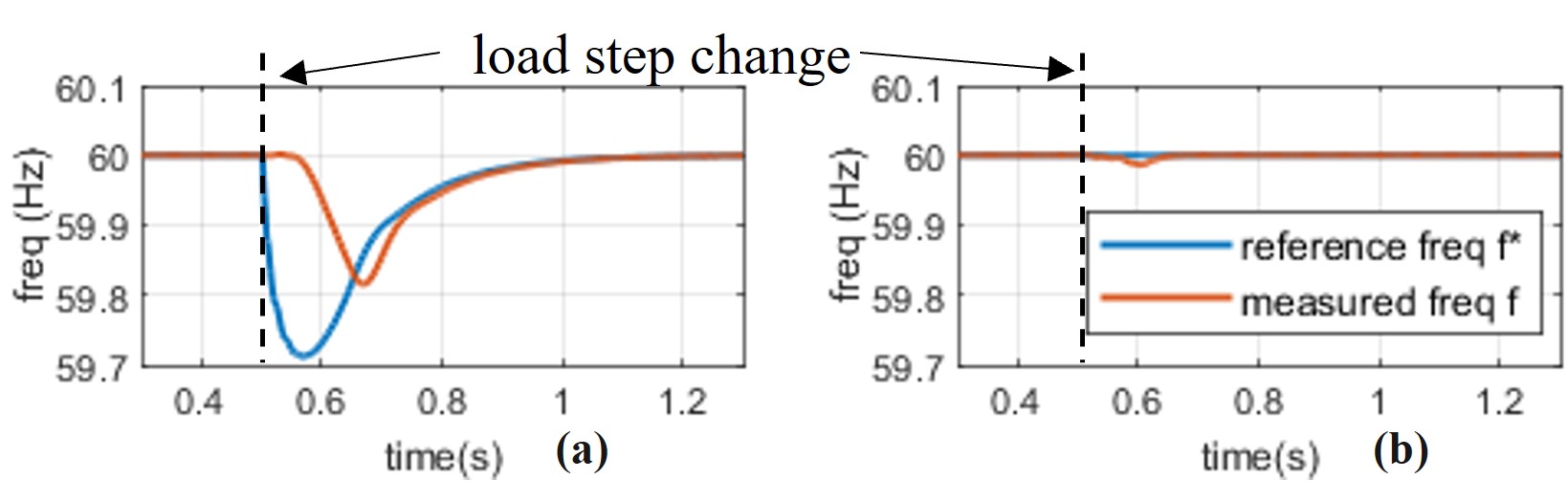}}
    \vspace{-15pt}
    \caption{Frequency response comparison (a) Conventional BESS control structure (b) Simplified, proposed BESS control structure.}
    \label{fig:f_ref}    
\end{figure}

In a MG with only one GFM resource, where control duty sharing among resources is unnecessary, the control structure can be significantly simplified. As illustrated in Fig. \ref{fig:control}(b), the secondary voltage controller is still required to regulate the voltage magnitude at PCC. However, in the absence of multiple GFM resources, the droop and secondary control become redundant, and the frequency reference can be directly sent to the inner voltage controller. This simplification not only reduces the complexity of the control system but also minimizes the hardware cost required for control. Most importantly, it enables faster dynamic response with reduced voltage and frequency transients. As shown in Fig. \ref{fig:f_ref}, the frequency reference no longer changes within  the inner control loop, resulting in significantly smaller frequency deviation and much quicker response times. We consider this simplification of the BESS frequency control.




\vspace{-5pt}
\subsection{UFLS Execution Devices}
\begin{figure}[t]
	\centerline{\includegraphics[width=0.5\textwidth]{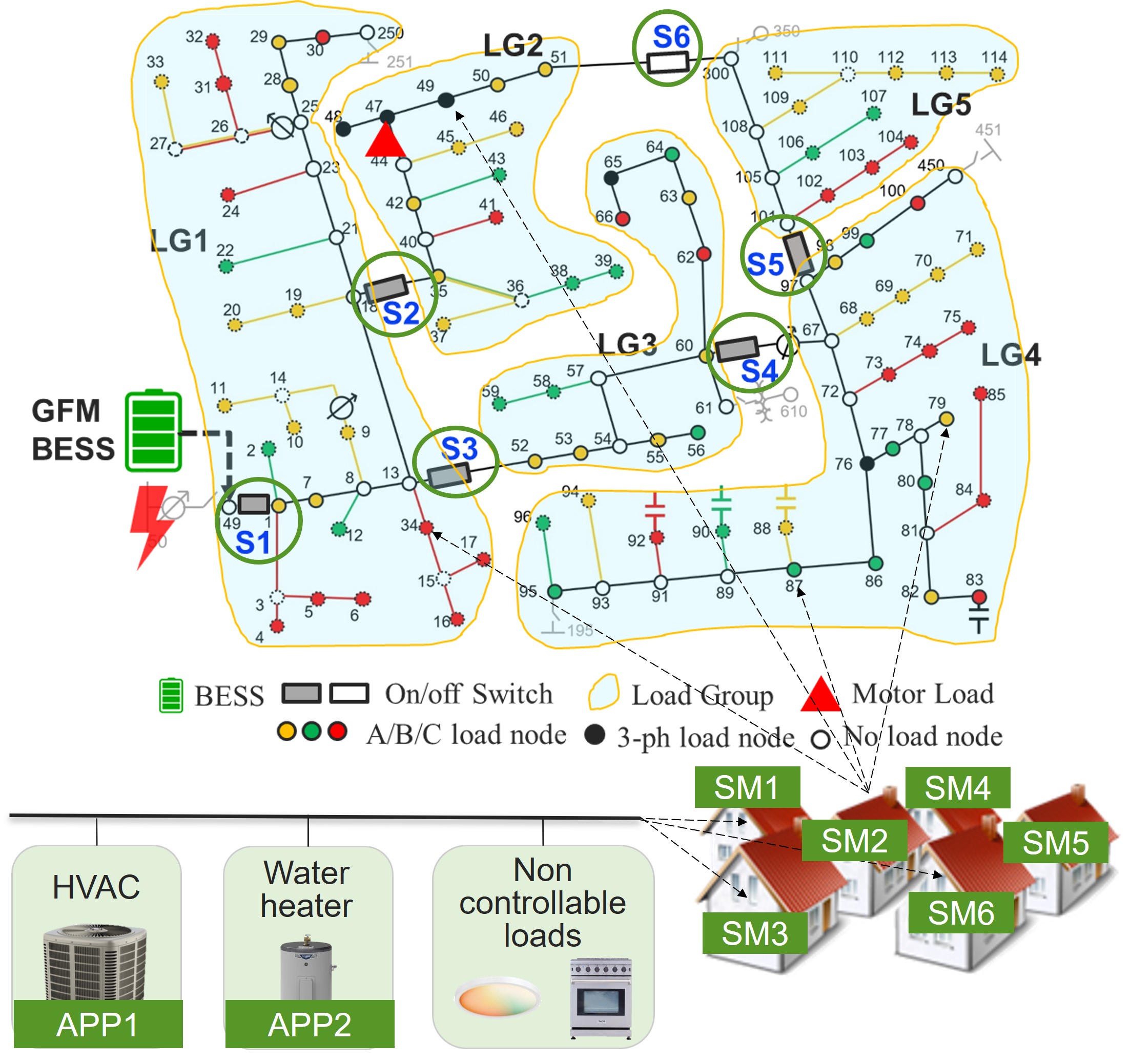}}
        \vspace{-10pt}
	\caption{Configuration of an islanded MG powered by a single GFM BESS.}
	\label{fig:MG}
\end{figure}

As depicted in Fig. \ref{fig:MG}, in an islanded MG, three controllable devices can be used to implement UFLS: sectionalizers (e.g., S1-S6), which can turn on/off an entire load group (LG); smart meters (e.g., SM1-SM6), capable of turning on/off an entire building/house; and controllable appliances (e.g., APP1 and APP2).

A summary of the advantages and disadvantages of conventional UFLS and the three proposed UFLS schemes has been provided in Table \ref{tab:SUM1}. In practice, controlling circuit breakers offers two primary advantages: simple implementation and reliable execution. 
Nonetheless, the UFLS approach suffers from a drawback: its non-selective, on/off-based load shedding often leads to over-shedding of loads or causes significant power surges during the recovery phase.

\vspace{+5pt}

On the other hand, the smart meter and appliance-based UFLS schemes offer the advantages of performing progressive, per-phase load shedding, allowing for selective, voluntary participation from the customers. Consequently, UFLS can not only provide power reserves but also reduce three-phase imbalances. The main drawback of these two methods is that integrating frequency control chips into a large amount of smart meters or appliances introduces additional costs. Additionally, well-designed autonomous UFLS coordination schemes are required for preventing synchronized turning off/on of controllable loads during UFLS execution or when restoring the shed loads.


\vspace{-5pt}
\subsection{Coordination between UFLS and the Embedded Controller}

\begin{figure}[t]
	 \vspace{-10pt}
  \centerline{\includegraphics[width=0.5\textwidth]{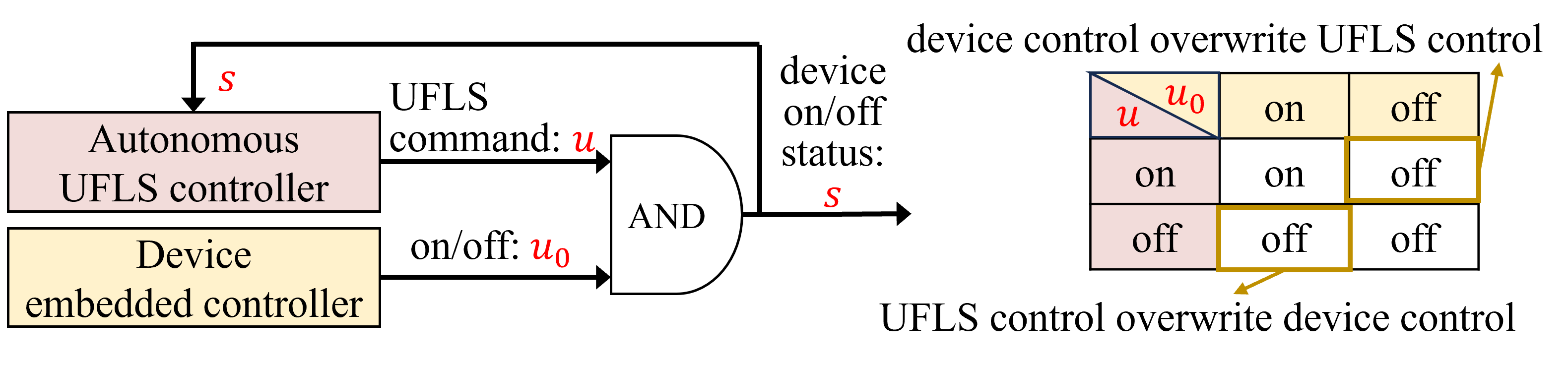}}
        \vspace{-10pt}
	\caption{Coordination between the UFLS control and the embedded, native control functions on a controllable load.}
	\label{fig:uisi}
\end{figure}

Fig. \ref{fig:uisi} demonstrates how the UFLS controller installed on an appliance actively monitors the appliance's on/off status ($s$) and executes switching on/off commands ($u$) in synchronization with the appliance's embedded function. This mechanism plays a crucial role in ensuring the safe operation of the appliance. For instance, when a water heater thermostat activates the appliance to heat up the water, a UFLS ``off'' command can turn off the water heater and reduce electricity consumption. Conversely, if the water heater is already in the ``off'' state, the UFLS ``on'' command cannot turn it on, because the temperature in the tank is already within the allowable range. This setting can prevent overheating of the water when executing UFLS commands.  On the other hand, if the water heater thermostat activates the appliance due to low water temperature, the water heater will not turn on until the UFLS ``off'' period is over, ensuring meeting UFLS requirements.


\vspace{-7pt}
\subsection{Device-level UFLS Triggering and Recovery Mechanism}
\begin{figure}[t]
	\vspace{-10pt}
 \centerline{\includegraphics[width=0.4\textwidth]{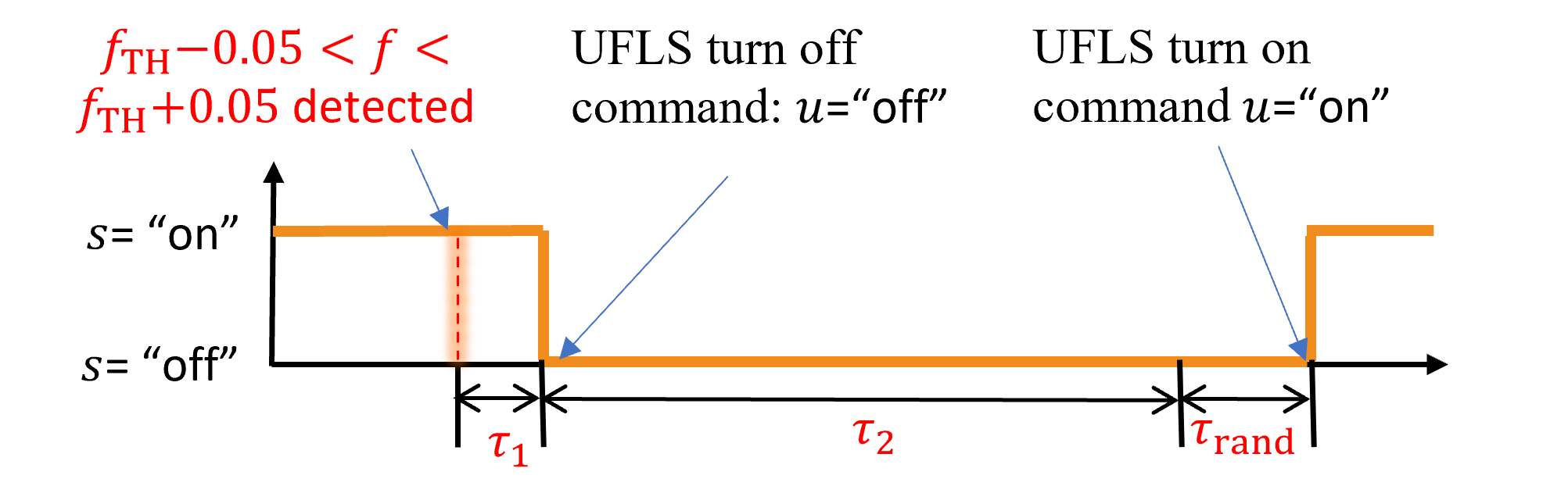}}
    \vspace{-10pt}
	\caption{An illustration of the UFLS response time delay, fixed recovery time, and random recovery delay during an UFLS event.}
    \vspace{+3pt}
	\label{fig:status}
\end{figure}

\begin{figure}[t]	
\centerline{\includegraphics[width=0.5\textwidth]{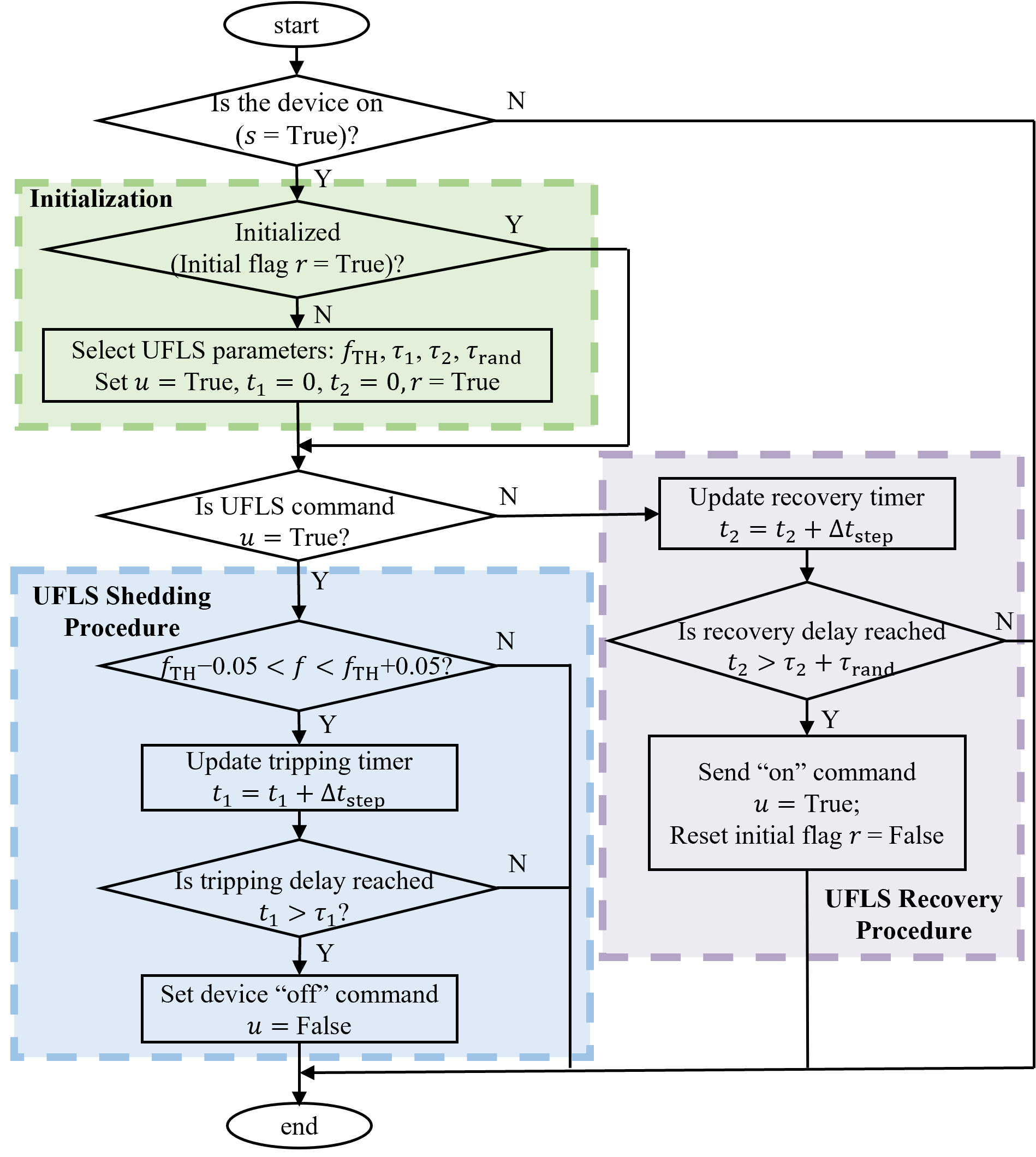}}
    \vspace{-10pt}
    \caption{Flow chart of the device-level UFLS scheme.}
    \label{fig:LS flow chart}
\end{figure}

As depicted in Fig. \ref{fig:status}, a UFLS device is characterized by four vital control parameters: the triggering frequency threshold ($f_{\mathrm{TH}}$), the tripping delay ($\tau_1$), fixed recovery delay ($\tau_2$), and the random recovery  delay ($\tau_{\mathrm{rand}}$). The flow chart of the detailed device-level UFLS control process is shown in Fig. \ref{fig:LS flow chart}.

Upon activation of the device ($s=\mathrm{True}$), the device-level UFLS controller will go through the initialization process, assigning initial values to each control parameter and resetting timers.
If the device command is ``on'' ($u=\mathrm{True}$), the system frequency $f$ is measured continuously at each time step ($\Delta t_{\mathrm{step}}$). If the measured frequency falls within the range of $f_{\mathrm{TH}}-0.05$ to $f_{\mathrm{TH}}+0.05$, a tripping timer ($t_1$) is initiated. Once the predefined turn-off delay $\tau_1$ is reached, the UFLS device will be deactivated and set back to ``off'' status. It is worth noting that the frequency deadband of 0.05 Hz is utilized to accommodate any frequency detection errors and is determined by the accuracy of the device's frequency sensor.

Once the UFLS device is deactivated, it remains in the ``off'' state until the designated recovery time ($\tau_2$) has elapsed. 
To ensure that all shed UFLS loads are not simultaneously turned on, the recovery timer ($t_2$) duration is determined by the sum of the fixed restoration time ($\tau_2$) and the individual restoration delay ($\tau_{\mathrm{rand}}$). 
Note that $\tau_2$ can be a constant for all UFLS devices  but $\tau_{\mathrm{rand}}$ is randomly selected by each UFLS device. For instance, if $\tau_2$ is set to 15 minutes, a UFLS device with a randomized delay of $\tau_{\mathrm{rand}}=1$ will be off for 16 minute, while another UFLS device with $\tau_{\mathrm{rand}}=2$ will be off for 17 minutes. This approach prevents synchronized activation and ensures a more distributed recovery of UFLS devices.

\subsubsection{Per-phase Triggering Frequency}
\begin{figure}[t]	
\vspace{-10pt}
\centerline{\includegraphics[width=0.43\textwidth]{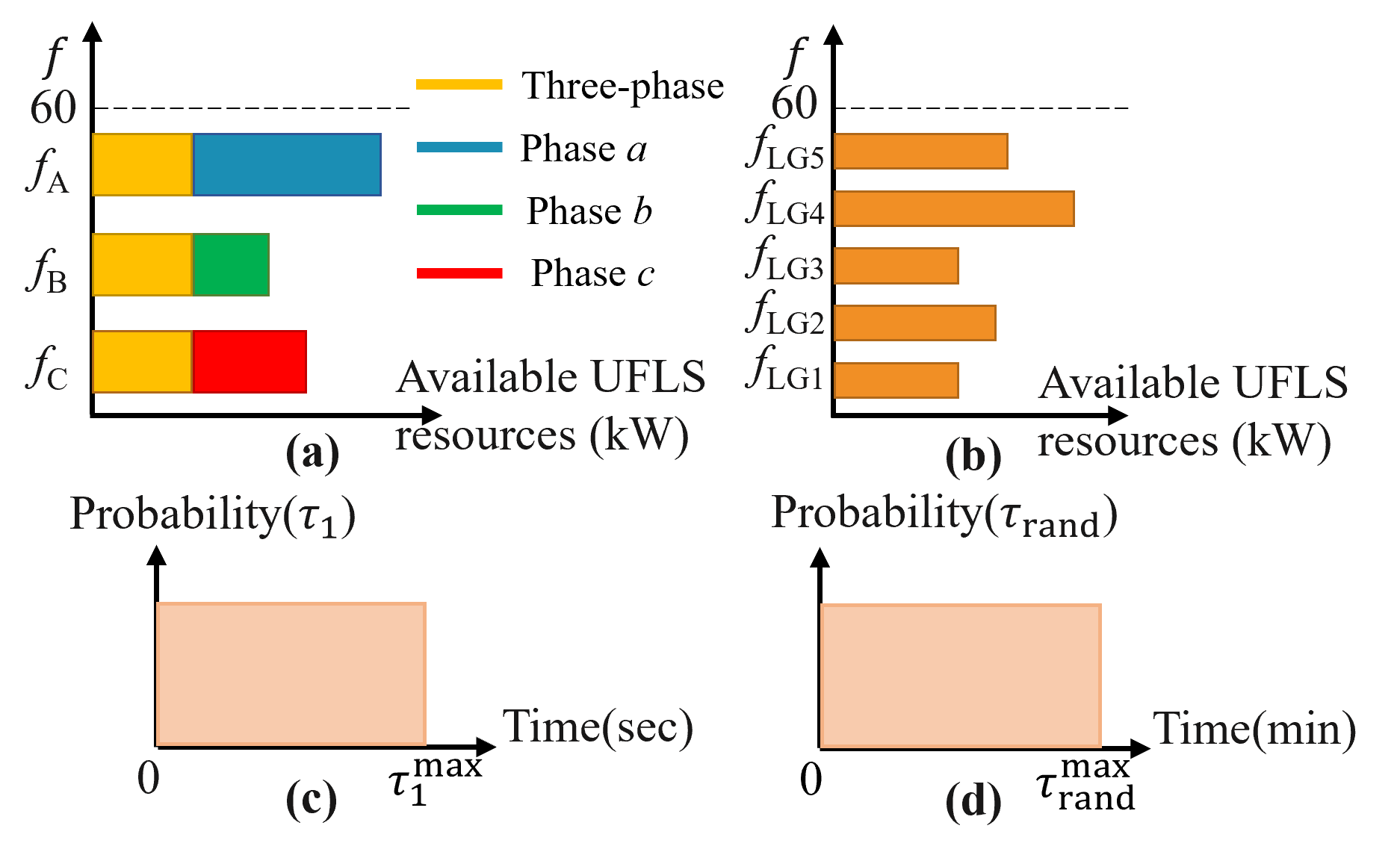}}
          \vspace{-10pt}
    	\caption{An illustration of the UFLS parameter setups (a) Phase-based triggering frequency setting; (b) load-group-based triggering frequency setting; (c) Tripping delay selection; (d) UFLS recovery delay selection.}
     \vspace{+5pt}
	\label{fig:LS parameters}
\end{figure}

One of the main contributions of this paper is that we introduce per-phase UFLS triggering frequency for the first time. As shown in Fig. \ref{fig:LS parameters}(a), for controllable appliances on phase $a$, $b$, and $c$, we assigned three unique triggering frequency thresholds. Thus, by lowering the system frequency to $f_\mathrm{A}$, $f_\mathrm{B}$, or $f_\mathrm{C}$, we can selectively trigger controllable resources based on phase $a$, $b$, or $c$.  Thus, if the phase $a$ power exceeds the power reserve threshold, we can shed phase $a$ load only. This per-phase UFLS is highly efficient when meeting PRR and can also be used to solve phase imbalance issues.  For a three-phase UFLS load, if $f$ falls into any of the three frequency detection bands, the UFLS mechanism will be triggered. This is because 3-phase loads can reduce the power consumption of all phases without exacerbating phase imbalance issues.

When sectionalizers are used for UFLS, an entire section of the feeder section is deactivated. Given that these feeder sections typically operate on a 3-phase basis, sectionalizer-based UFLS schemes inherently lack the capacity for load shedding on a per-phase basis. Consequently, to enable a sequential shedding process using sectionalizers, individual frequency setpoints are designated to each sectionalizer in a descending order, as depicted in Fig. \ref{fig:LS parameters}(b).

Conventional UFLS frequency thresholds are set to be 0.25 to 0.3 Hz apart so two UFLS groups will not be triggered simultaneously \cite{li2019continuous}. In an islanded MG, the distance between the frequency setpoints depends on the maximum frequency deviations caused by typical step load changes. For example, if frequency deviations can be limited within 0.1 Hz, then we can select the frequency setpoints for phase $a$, $b$, and $c$ loads to be: $f_\mathrm{A}$=59.85 Hz, $f_\mathrm{B}$=59.55 Hz, $f_\mathrm{C}$=59.25 Hz, respectively. The frequency setpoints used in sectionalizer-based method should have the same frequency difference.
In the sectionalizer-based method, the frequency setpoints can maintain this identical frequency difference, reflecting the same precision.

\subsubsection{UFLS Tripping Delay}
Once an UFLS device detects that the system frequency falls within $f_{\mathrm{TH}} \pm 0.05$ Hz, it will select $\tau_{1}$ from a uniform distribution ($\sim U(0, \tau_1^{\mathrm{max}})$) by:

\begin{equation}\label{ttrip}
\vspace{-3pt}
\tau_{1} \sim U(0, \tau_1^{\mathrm{max}}), \quad \text{for} \quad i = 1, 2, \ldots, n
\vspace{-3pt}
 \end{equation}
where $\tau_1^{\mathrm{max}}$ is the maximum tripping delay, $n$ is the number of participant device.  

Tripping delay is the key parameter for progressively shedding loads to meet PRR. As shown in Fig. \ref{fig:LS parameters}(c), the upper bound of the range $\tau_1^{\mathrm{max}}$ is contingent on the lowest frequency setpoint $f_\mathrm{min}=min(f_\mathrm{A}~f_\mathrm{B}~f_\mathrm{C})$. 
NERC reliability standard 
PRC-024-2 \cite{settings2018nerc} and IEEE standard 1547-2018\cite{standard1547} outline the frequency ride-through performance characteristic for generator’s protective relay settings. To avoid tripping DERs (e.g., rooftop PV systems), $\tau_1^{\mathrm{max}}$  should be set to a value less than the low-frequency duration associated with the lowest frequency setpoint, as specified in the standards:

\begin{equation}\label{ttripmax}
\tau_1^{\mathrm{max}}  < 10^{(1.7373*f_\mathrm{min}-100.116)}
\end{equation}

\subsubsection{UFLS Recovery Delay}
After a UFLS device is switched off, it will remain off for a duration of $\tau_{2} + \tau_{\mathrm{rand}}$. The fixed recovery delay, $\tau_2$, can be set to match the duration of a typical DR event, such as 15 minutes. As illustrated in Fig. \ref{fig:LS parameters}(d), to prevent synchronized turning-on of all UFLS devices, a random time delay $\tau_{\mathrm{rand}}$ is chosen from a uniform distribution:
\begin{equation}\label{tback}
\tau_{\mathrm{rand}} \sim U(0, \tau_{\mathrm{rand}}^{\mathrm{max}}), \quad \text{for} \quad i = 1, 2, \ldots, n
\vspace{-3pt}
\end{equation}
where $\tau_{\mathrm{rand}}^{\mathrm{max}}$ is the maximum individual recovery delay.  


\begin{figure*}[htb]	
    \vspace{-10pt}
    \centerline{\includegraphics[width=\textwidth]{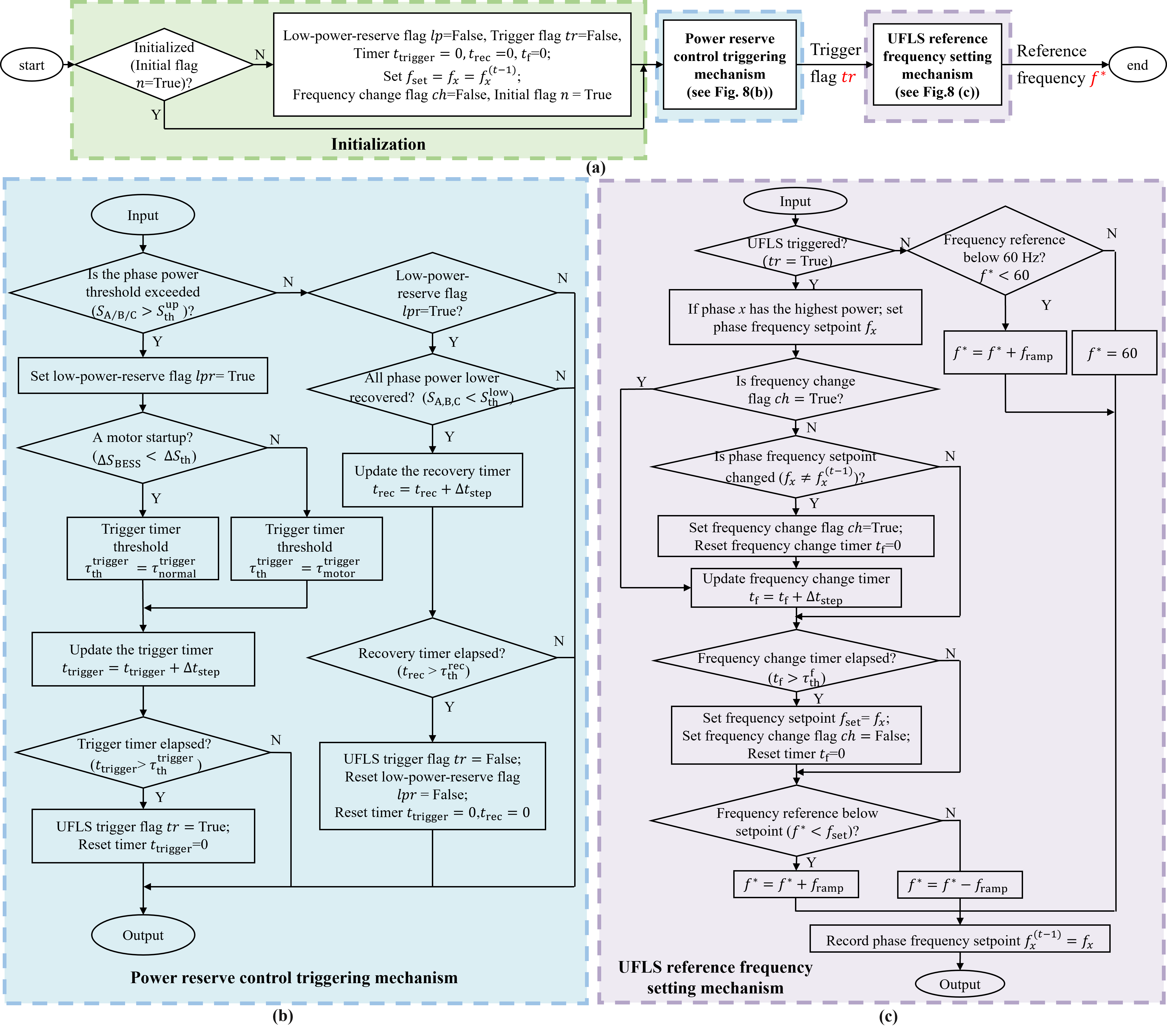}}
    \vspace{-14pt}
	\caption{Flow chart of BESS frequency regulation mechanism for meeting PRR (a) BESS frequency regulation mechanism; (b) Power reserve control triggering mechanism; (c) UFLS reference frequency  setting mechanism.}
	\label{fig:BESS flow chart}
 \vspace{-10pt}
\end{figure*} 


\vspace{-5pt}
\subsection{Frequency Regulation Mechanism for Meeting PRR}
In this paper, we assume that there is only one GFM BESS in the islanded MG, and it is solely responsible for system frequency regulation. This allows us to use the simplified frequency control structure depicted in Fig. \ref{fig:control}(b) to modulate the system frequency as the UFLS control signal. However, it's important to acknowledge that if there are multiple GFM resources regulating frequency, coordination among the GFM devices must be considered. We recognize this as a potential area for future research.

The BESS frequency control mechanism displayed in Fig. \ref{fig:BESS flow chart} comprises three essential processes: initialization, power reserve control triggering mechanism, and UFLS reference frequency setting mechanism. The initialization mechanism is used for assigning initial values for all controllable parameters and resetting timers.

\subsubsection{Power Reserve Control Triggering Mechanism}
In normal operation, the BESS monitors the per-phase upper power reserve threshold ($S_{\mathrm{th}}^{\mathrm{up}}$) calculated by:
\begin{equation}\label{Sth}
\vspace{-3pt}
S_{\mathrm{th}}^{\mathrm{up}}=1-S_{\mathrm{PR}}
\vspace{-3pt}
\end{equation}
where $S_{\mathrm{PR}}$ is the PRR. Note that the per unit values are used for all $S$ values.

\textbf{Low-power-reserve Detection:} If any of the three-phase power reserve limits exceed the threshold, i.e., $S_{\mathrm{A/B/C}}>S_{\mathrm{th}}^{\mathrm{up}}$, the low-power-reserve flag ($lpr$) is set to ``True'' and the BESS activates the triggering timer ($t_{\mathrm{trigger}}$).  The time delay ($\tau^{\mathrm{trigger}}_{\mathrm{th}}$) is used to determine when to activate the UFLS trigger flag $tr$ so that frequency setpoint $f^*$ can be dropped to corresponding UFLS triggering threshold. Then, the frequency change can be detected by the frequency chips on sectionalizers, smart meters, or appliance controllers to activate autonomous device level UFLS following the device level UFLS mechanism shown in Fig. \ref{fig:LS flow chart}.  

\textbf{Avoid Power Surges:} To prevent unnecessary responses to short-lived power surges (such as motor startups), we compare the Rate of Change of Load Power (ROCOLP) ($\Delta S_{\mathrm{BESS}}$) with the ROCOLP threshold ($\Delta S_{\mathrm{th}}$). If the change in the BESS capacity, denoted as $\Delta S_{\mathrm{BESS}}$, is less than the threshold $\Delta S_{\mathrm{th}}$, the trigger timer ($t_
{\mathrm{trigger}}$) will be set with a threshold of $\tau_{\mathrm{th}}^{\mathrm{trigger}}$. Typically, $\tau_{\mathrm{th}}^{\mathrm{trigger}}=\tau_{\mathrm{normal}}^{\mathrm{trigger}}$ can be chosen to be 1 to 3 cycles (approximately 0.02-0.06 seconds) to allow the measuring of a steady-state power value \cite{setiabudy2013analysis}. If the change in the BESS capacity, denoted as $\Delta S_{\mathrm{BESS}}$, exceeds or is equal to the threshold $\Delta S_{\mathrm{th}}$, the parameter $t_{\mathrm{trigger}}$ will be set with a threshold of $\tau_{\mathrm{th}}^{\mathrm{trigger}}=\tau_{\mathrm{motor}}^{\mathrm{trigger}}$. In this scenario, the UFLS trigger flag $tr$ will only activate if the phase power $S_{\mathrm{A/B/C}}$ remains above $S_{\mathrm{th}}^{\mathrm{up}}$ for a duration longer than $\tau_{\mathrm{motor}}^{\mathrm{trigger}}$. This designed triggering delay will ensure that short-lived power spikes do not mistakenly trigger the UFLS process.

\textbf{Power Reserve Recovered:} The UFLS process continues until all phase powers fall below the specified low power threshold, denoted as $S_{\mathrm{A,B,C}}<S_{\mathrm{th}}^{\mathrm{low}}$. This condition indicates that the power reserve has been restored, and the previously shed loads can now be safely restored. To prevent power oscillation caused by frequent activation of UFLS, it is advisable to set the value of $S_{\mathrm{th}}^{\mathrm{low}}$ approximately 0.05 to 0.1 p.u. lower than $S_{\mathrm{th}}^{\mathrm{up}}$.  The UFLS recovery timer ($t_{\mathrm{rec}}$) can be set with a threshold of $\tau_{\mathrm{th}}^{\mathrm{rec}}$. Once this threshold is reached, the UFLS trigger flag will be reset, and all parameters will undergo re-initialization.

\subsubsection{UFLS Frequency Reference Selection Mechanism}
Once the UFLS trigger flag $tr$ is set to be ``True'', the GFM controller will select $f^*$. Note that $f^*$ is the frequency reference signal sent to the BESS GFM controller shown in Fig. \ref{fig:control}(b) for adjusting the system frequency to its setpoint ($f_{\mathrm{set}}$). 

\textbf{Frequency Ramp Limits}: In normal operation, the Rate of change of frequency (ROCOF) limit is typically set to a value varying from 0.1 Hz/s to 0.5 Hz/s \cite{broderick2018rate,chown2017system} for ensuring that the frequency remains stable within a narrow range around its nominal value. Thus, when the ramp-up and ramp-down slopes of $f^*$ should be set as $f_\mathrm{ramp}=(0.5$ Hz/s) $\times \Delta t_\mathrm{step}$ to comply with the normal operation requirements, where $\Delta t_\mathrm{step}$ is the execution time step of GFM BESS controller.

\textbf{UFLS Phase Selection and Transition}: Once the UFLS is triggered ($tr=\mathrm{True}$), the BESS controller will initially identify the phase $x$ ($x\in {\mathrm{A}, \mathrm{B}, \mathrm{C}}$) with the highest power that exceeds the upper power threshold ($S_{\mathrm{th}}^\mathrm{up}$), specifically satisfying $S_x>S_{\mathrm{th}}^\mathrm{up}$ and $S_x = \max(S_\mathrm{A}, S_\mathrm{B}, S_\mathrm{C})$. Subsequently, the controller will compare the phase frequency $f_x$ at time $t$ with that of $t-1$, i.e., assessing whether $f_x^{(t)}$ is equal to $f_x^{(t-1)}$, to determine if a phase shift for UFLS is necessary. For instance, assuming phase $b$ has the highest power exceeding the threshold, causing $f_{\mathrm{set}} = f_\mathrm{B}$ at time $t-1$. If at time $t$, phase $a$ exhibits the highest power, a phase shift occurs, leading to a change in $f_{\mathrm{set}}$ from $f_\mathrm{B}$ to $f_\mathrm{A}$. 

However, it's important to note that the BESS will not immediately make $f_{\mathrm{set}}=f_\mathrm{A}$ once the phase shift is detected. Instead, a frequency setpoint change timer ($t_\mathrm{f}$) will be activated. This ensures that a frequency setpoint $f_{\mathrm{set}}$ will only be altered after a timer delay ($\tau _\mathrm{th}^\mathrm{f}$), thus avoiding responding to short-lived power fluctuations.

\textbf{Frequency Reference Selection}:
Once the $f_\mathrm{set}$ is determined, the GFM BESS frequency reference $f^*$ will be adjusted gradually to converge towards the frequency setpoint $f_\mathrm{set}$ at a controlled ramp rate of $f_\mathrm{ramp}$. After the UFLS event concludes, $f^*$ will be reverted back to its original value of 60 Hz, as illustrated in Fig. \ref{fig:BESS flow chart}.

\subsubsection{An Illustration of the Frequency Regulation Process}
\begin{figure}[t]
	\vspace{-10pt}
 \centering	\centerline{\includegraphics[width=0.48\textwidth]{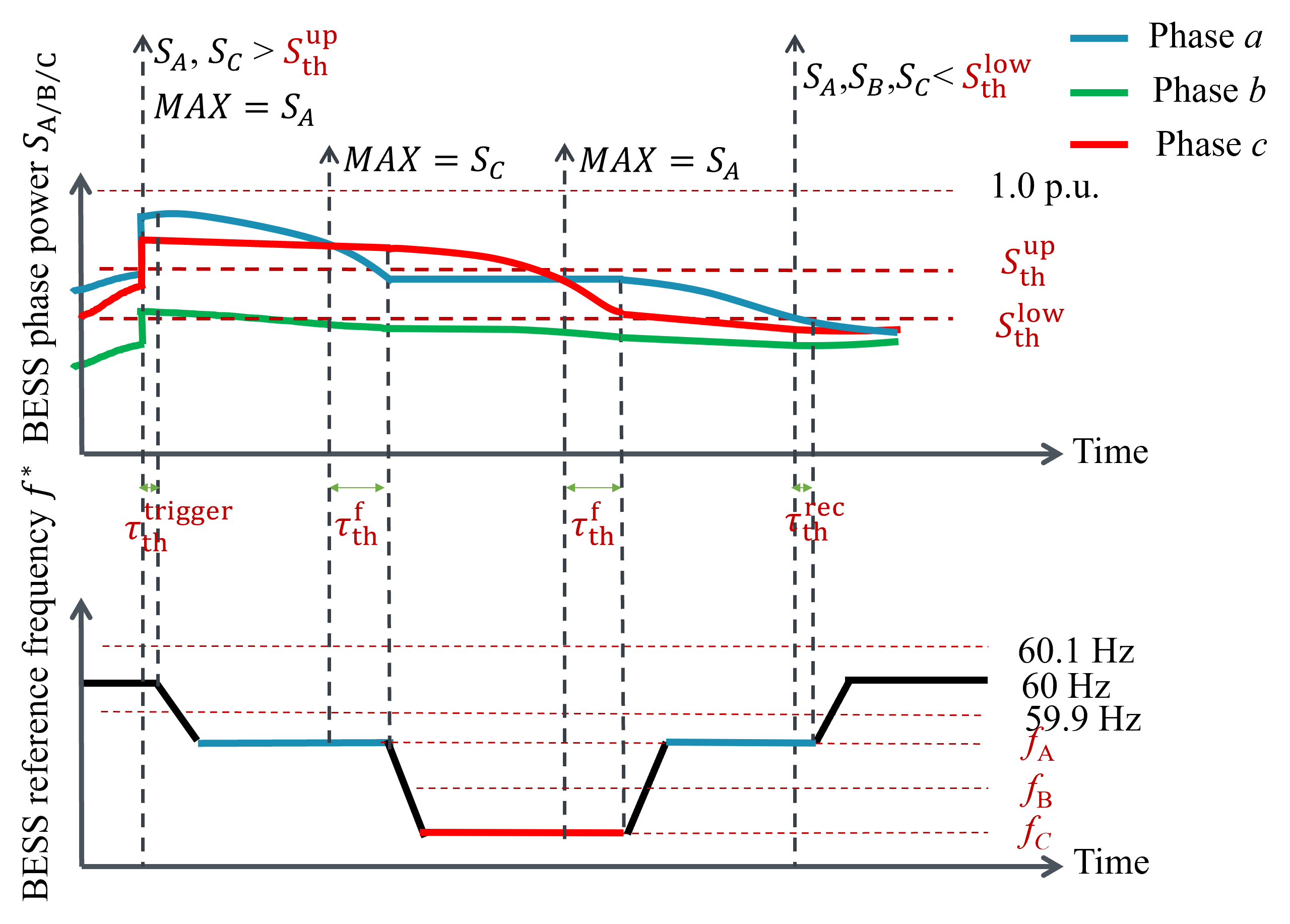}}
     \vspace{-10pt}
	\caption{An illustration of the BESS 3-phase power output variations and corresponding BESS reference frequency adjustment for triggering a UFLS event to meet PRR.}
	\label{fig:state}
\end{figure}

In Fig. \ref{fig:state}, the upper diagram displays the BESS phase apparent power, while the lower diagram illustrates the changes in the GFM BESS reference frequency $f^*$. Initially, both phase $a$ and phase $c$ exhibit power outputs that surpass the power reserve threshold, denoted as $S_\mathrm{A}, S_\mathrm{C}>S_{\mathrm{th}}^{\mathrm{up}}$. Consequently, the UFLS is triggered following a time delay of $\tau_{\mathrm{th}}^{\mathrm{trigger}}$.

Due to the highest power output in phase $a$, the reference frequency is set as $f^*=f_\mathrm{A}$. This initiates the load shedding process for devices connected to phase $a$, leading to a reduction in the BESS power output in this phase. Subsequently, when the phase with the highest power output shifts from $a$ to $c$, the reference frequency adjusts accordingly after a time delay of $\tau_{\mathrm{th}}^{\mathrm{f}}$, ensuring a smooth transition.
The updated frequency setpoint is then set to $f^*=f_\mathrm{C}$, triggering the shedding of phase $c$ loads. This iterative process continues until the power in all phases falls below $S_{\mathrm{th}}^{\mathrm{low}}$, causing the frequency to ramp up back to 60 Hz.

By this procedural sequence, the proposed UFLS scheme effectively orchestrates load shedding based on phase power thresholds, ultimately achieving 3-phase balance regulation.

\section{Simulation Results}
To assess the performance of the proposed UFLS scheme, a MG testbed based on the IEEE 123-bus topology was developed on the OPAL-RT platform. As depicted in Fig. \ref{fig:MG}, a 3-MVA GFM BESS with the topology shown in Fig. \ref{fig:control} is connected to Node 149 at the feeder head. Note the BESS serves as the sole GFM resource within the MG. To evaluate the UFLS during a motor start-up event, a 400 kVA three-phase motor load is connected to Node 47 in LG2. The parameters used in the simulation are listed in Table~\ref{parameter}.
The GFM BESS model is simulated at the microsecond level in eMEGASIM. The ZIP load models, motor load model, and the unbalanced IEEE 123-bus network model are used to emulate the operation of realistic distribution feeders, which are simulated in ePHASORSIM at the milisecond level. For more detailed BESS model parameters and simulation setups, please refer to \cite{xu2022novel, paduani2022novel}.

A one-hour test involving the MG's blackstart process has been conducted to verify the effectiveness of the proposed UFLS approaches for managing MG power reserves. Load groups LG1, LG2, LG3, LG4, and LG5 will be sequentially powered up by activating switches S1, S2, S3, S4, and S5 at specific times: $t=0.3, 50, 100, 150, 200$ seconds.  Note that S6 remains deactivated to prevent the formation of loops within the MG. A set of load profiles capable of triggering UFLS events has been chosen. We focus on comparing two UFLS mechanisms: the sectionalizer-based and the appliance-based mechanisms. As the smart-meter based method produces outcomes very similar to those of the appliance-based method, we omit the comparison due to space constraints.

\begin{table}[!t] 
  \centering
  \vspace{-10pt}
  \caption{UFLS Control Parameters}\label{parameter}
  \vspace{-5pt}
  \begin{tabularx}{\columnwidth}{p{5.4cm}||c}
    \hline
    \hline
     Triggering frequency ($[f_\mathrm{A}~f_\mathrm{B}~f_\mathrm{C}]$) & [59.85 59.55 59.25] Hz\\
     Max. tripping delay  ($\tau_\mathrm{1}^\mathrm{max}$)  & 10 $s$\\
     Fixed recovery delay ($\tau_\mathrm{2}$) & 15 minutes\\
     Max. individual recovery delay ($\tau_\mathrm{rand}^\mathrm{max}$) & 3 minutes\\
     Device controller simulation time step ($\Delta t_\mathrm{step}$)  & 10 $ms$\\
     BESS controller simulation time step ($\Delta t_\mathrm{step}$)  & 100 $\mu s$\\
     Upper power  ($S_\mathrm{th}^\mathrm{up}$) & 0.9 p.u. \\
     Low power  ($S_\mathrm{th}^\mathrm{low}$) & 0.87 p.u. \\
     ROCOLP  ($\Delta S_\mathrm{th}$) & 0.5 p.u. \\
     Triggering timers ($\tau_\mathrm{normal}^\mathrm{trigger}$ and $\tau_\mathrm{motor}^\mathrm{trigger}$) & 0.02 $s$, 10 $s$\\
     Recovery timer  ($\tau_\mathrm{th}^\mathrm{rec}$) &  0.02 $s$\\
     Frequency change timer  ($\tau_\mathrm{th}^\mathrm{f}$) & 1 $s$\\
     Frequency ramp rate ($f_\mathrm{ramp}$) & 0.5 Hz/s\\
     \hline
     \hline
   \end{tabularx}
\end{table}

\vspace{-5pt}
\subsection{Case 1: Sectionalizer-based UFLS}
In case 1, we assess the performance of power reserve management by shedding LGs via sectionalizers. Note that in this case, UFLS devices can be either sectionalizers or breakers. Different from appliance or smart meter based approaches, the switching on/off of sectionalizers will result in an entire 3-phase feeder section to loss power supply.

We configure the trigger frequencies for LG1, LG2, LG3, LG4, and LG5 as 59.85, 59.55, 59.25, 58.95, and 58.65 Hz, respectively. The tripping delay ($\tau_\mathrm{1}$) is 0.02 s. The fixed restoration delay is $\tau_\mathrm{2}=$15 minutes and there is no random recovery delay, i.e., $\tau_\mathrm{rand}=0$.

\begin{figure}[t]
	\centering
    \vspace{-10pt}
	\centerline{\includegraphics[width=0.49\textwidth]{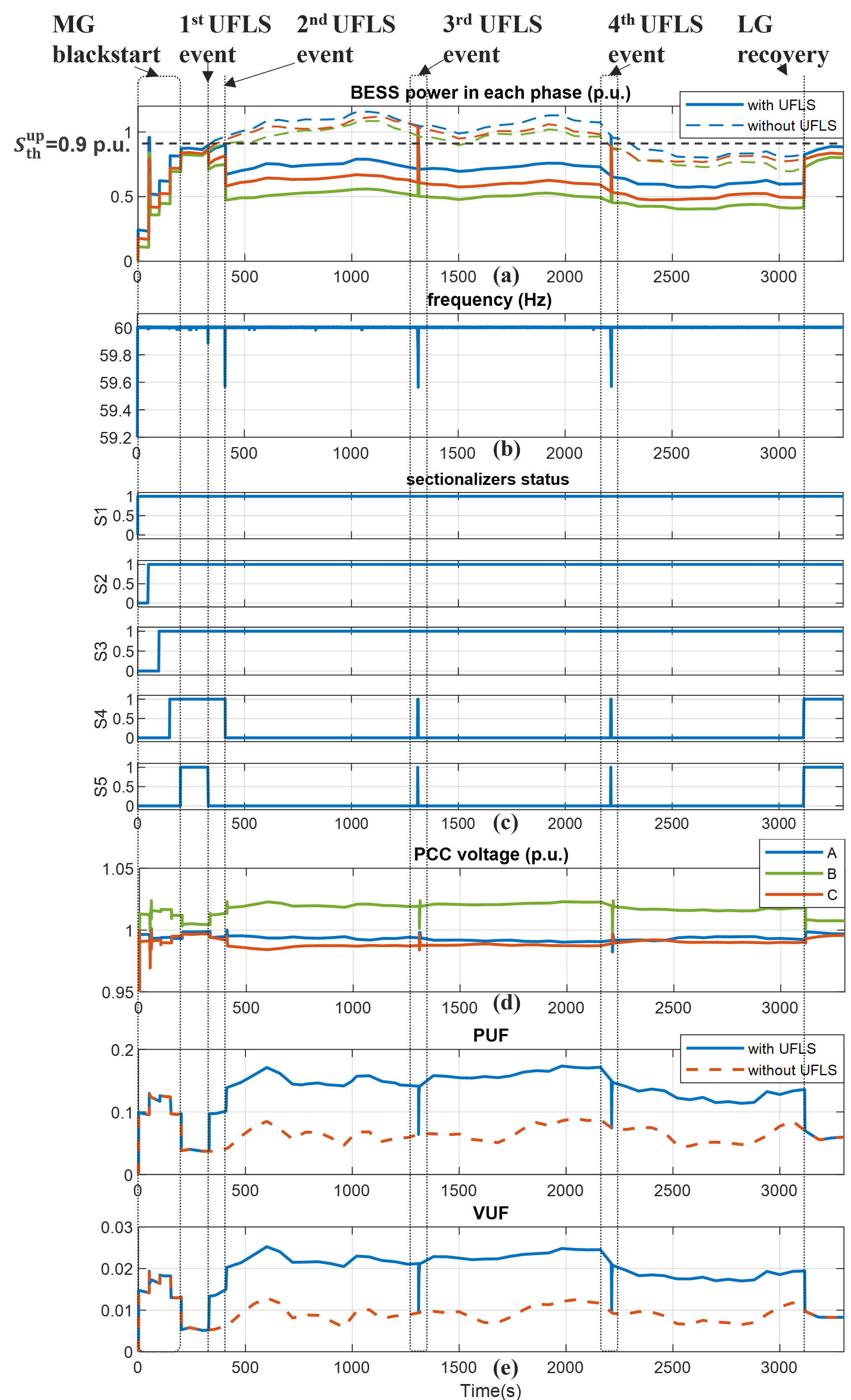}}
    \vspace{-10pt}
	\caption{Simulation results of the sectionalizers-based UFLS case. (a) BESS 3-phase power outputs; (b) System frequency; (c) Sectionalizer on/off status; (d) PCC voltage; (e) Phase-imbalance metrics: PUF and VUF.}
	\label{fig:re_CB}
\end{figure}

The dynamic responses of the phase power output of the BESS, system frequency, sectionalizer status, PCC voltage, and unbalance factors during this one-hour period are shown in Fig. \ref{fig:re_CB}.
As depicted in Fig. \ref{fig:re_CB}(a), the BESS phase power exceeds 1.0 p.u. at 450 s and persists until 2200 s when the power reserve control is not implemented. This scenario poses a significant risk of a complete shutdown of the MG. On the other hand, if the power reserve control is implemented, a low-power-reserve flag ($lpr$) will be sent to the BESS controller, indicating that the power reserve upper limit ($S_\mathrm{th}^{\mathrm{up}}=0.9$) has been exceeded and the BESS frequency regulation scheme will be triggered to shed loads and restore the power reserve.

From Fig. \ref{fig:re_CB}, four UFLS events can be observed during the one-hour MG blackstart process for meeting PRR. Table \ref{tab:UFLS events} summarizes the cause and actions taken during the four UFLS events.
The $1^\mathrm{st}$ UFLS event is triggered at approximately $t=330$ s, resulting in the shedding of LG5. As the loads continue to increase, a $2^\mathrm{nd}$ UFLS event is soon triggered at $t=410$ s, causing the shedding of LG4. The loads then remain under 0.8 p.u. after that.
After the 15-minute recovery delay elapses, LG4 and LG5 are switched on again. The sudden increase of power consumption immediately triggers the $3^\mathrm{rd}$ UFLS event at $t=1310$ s. The $4^\mathrm{th}$ UFLS occurs at $t=2210$ s, and is also caused by the re-connection of LG4 and LG5. Finally, at $t=3110$ s, the power reserve can be satisfied after LG4 and LG5 are reconnected.  

\begin{table}[]
    \centering
    \caption{UFLS events in case 1}
    \vspace{-5pt}
    \begin{tabular}{|c|c|C{3cm}|C{2.6cm}|}
        \hline
        Events & Time & Cause & Actions \\
        \hline
        \hline
        $1^\mathrm{st}$ & 330 s & High power demand & shed LG5 \\
        \hline
        $2^\mathrm{nd}$ & 410 s & High power demand & shed LG4 \\
        \hline
        $3^\mathrm{rd}$ & 1310 s & Restore LG4\&LG5 & Shed LG5 then LG4 \\
        \hline
        $4^\mathrm{th}$ & 2210 s & Restore LG4\&LG5 & Shed LG5 then LG4\\
        \hline      
    \end{tabular}
    \label{tab:UFLS events}
\end{table}

To evaluate phase imbalance, we use two metrics: the power unbalance factor (PUF) and voltage unbalance factor (VUF) defined in standards \cite{VUF_definition, xu2022novel}. As shown in Fig. \ref{fig:re_CB} (c), there is a violation in both PUF and VUF during the load shedding process. This shows that one of the drawbacks of the sectionalizer-based UFLS method is that it is incapable of regulating phase imbalance and can inadvertently exacerbate imbalance issues.  

\begin{figure}[t]
	\centering
    \vspace{-10pt}
	\centerline{\includegraphics[width=0.48\textwidth]{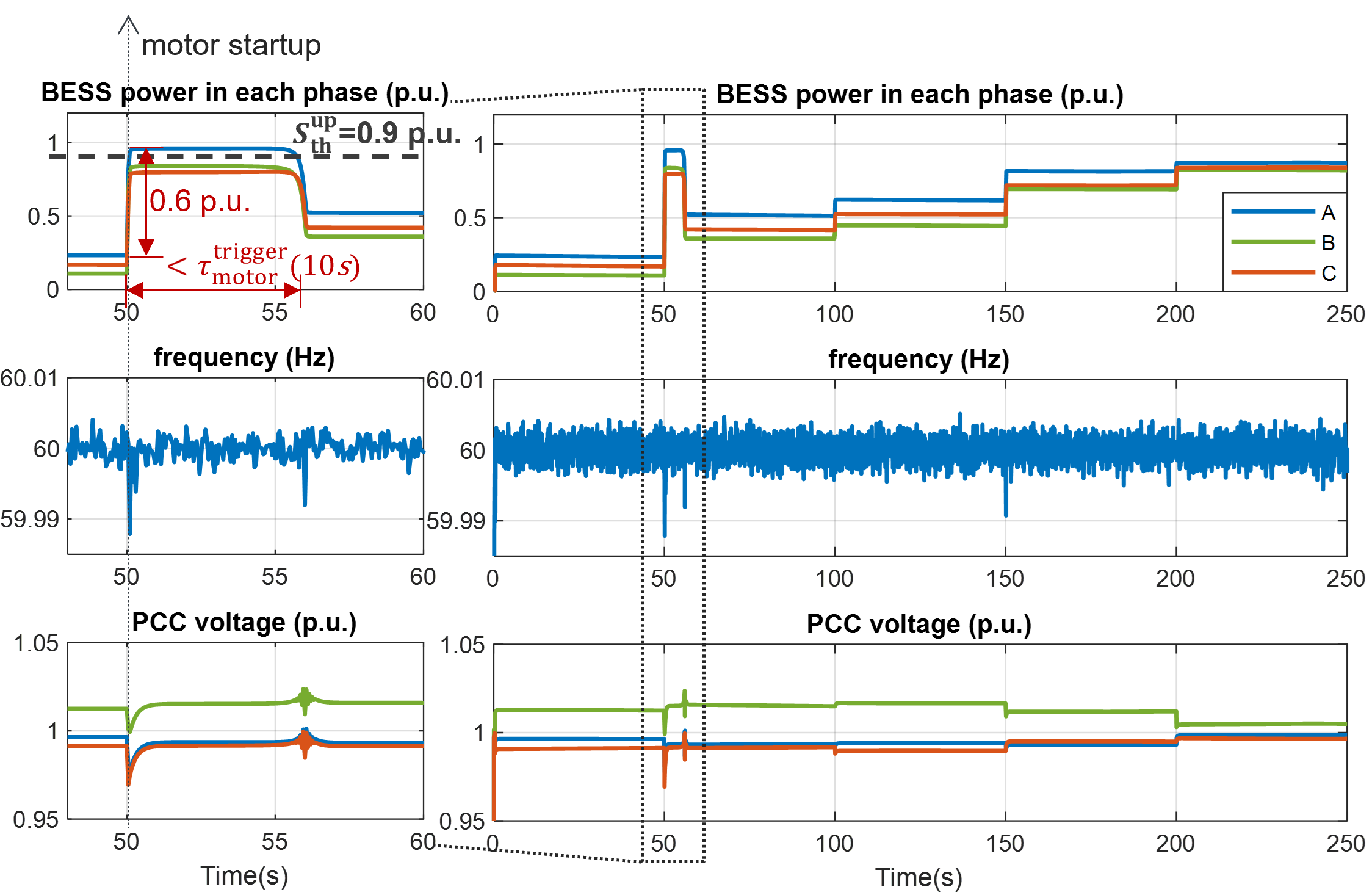}}
    \vspace{-10pt}
	\caption{PCC voltage, system frequency, and BESS 3-phase power outputs during the blackstart period.}
	\label{fig:re_motor}
\end{figure}

As shown in the zoom-in plot of the blackstart process (0 to 250 s) in Fig. \ref{fig:re_motor}, a three-phase motor connected to LG2 is energized at $t=50$ s.
To avoid undesirable response to short-lived power surges caused by motor start-up or cold-load pickup processes, the ROCOLP threshold is set to $\Delta S_\mathrm{th}=$ 0.5 p.u. From the results, we can see that although the BESS power output in phase $a$ is larger than 0.9 p.u. during 50-56 s, the step change of BESS power ($\Delta S_\mathrm{BESS}$) is larger than the threshold $\Delta S_\mathrm{th}=$ 0.5 p.u., hence the UFLS is not triggered. In this case, the UFLS is successfully avoided because the duration of the high power period is less than the motor triggered timer threshold $\tau_\mathrm{motor}^\mathrm{trigger}=10$ s. 

\begin{figure}[htb]
	\centering
	\centerline{\includegraphics[width=0.49\textwidth]{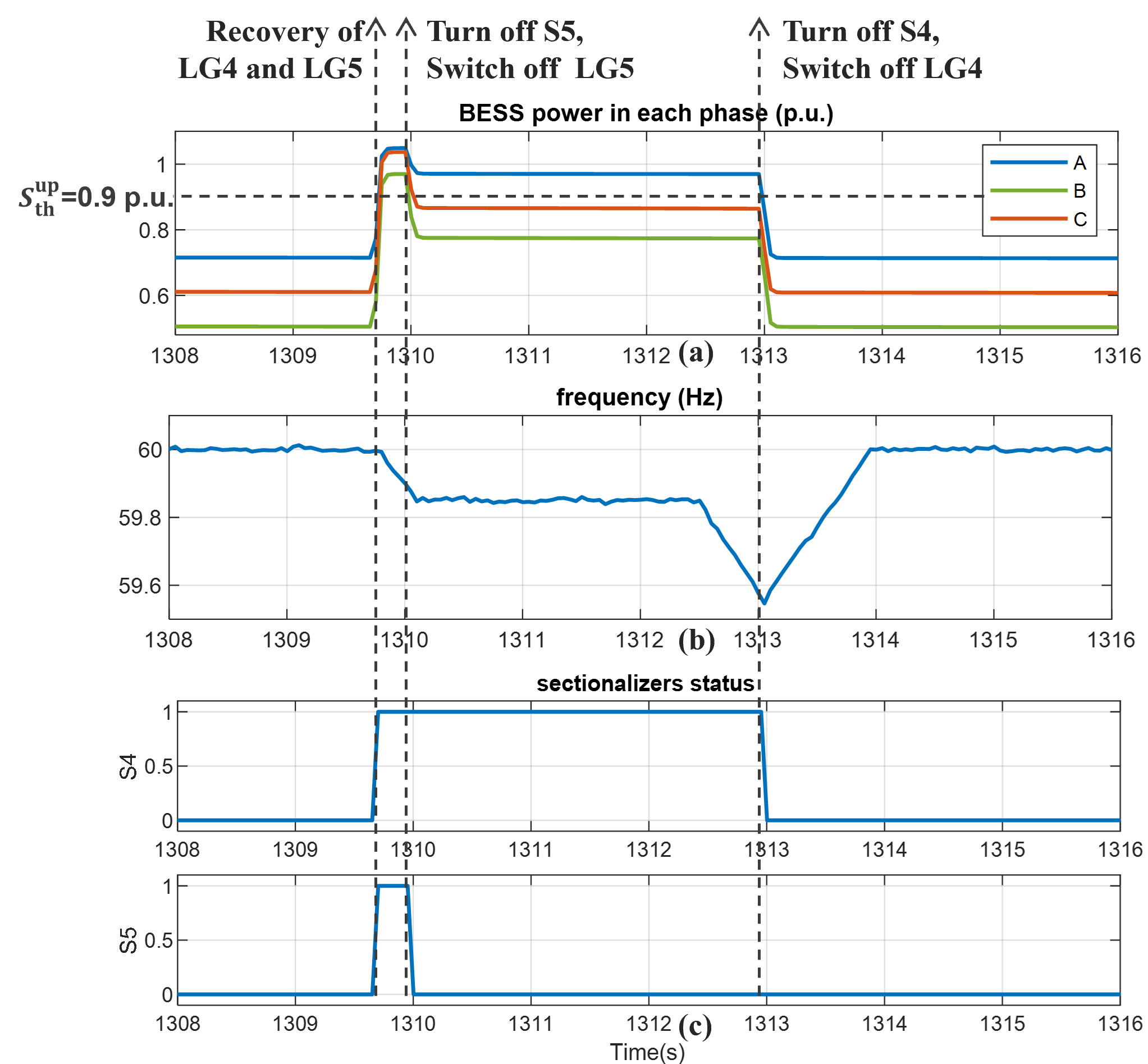}}
  \vspace{-10pt}
    \caption{BESS 3-phase power outputs, system frequency, sectionalizers S4 and S5 on/of status during the third UFLS event.}
	\label{fig:re_CB_in}
\end{figure}

Figure \ref{fig:re_CB_in} displays a complete UFLS event occuring during seconds 1309 to 1314. Initially, LG4 and LG5 are shed at the $1^\mathrm{st}$ and $2^\mathrm{nd}$ UFLS event, as shown in Fig. \ref{fig:re_CB}. After the recovery time of 15 minutes, shed loads in LG4 and LG5 are reconnected at 1309.8 s. Subsequently, since the BESS power output exceeds the upper power threshold ($S_\mathrm{th}^{\mathrm{up}}=0.9$ p.u), another UFLS event is triggered at $t=1310$ s. The system frequency starts to decline with a rate of 0.5 Hz/s until it reaches 59.85 Hz, after which, switch S5 is turned off.
However, since phase $a$ power remains over 0.87 p.u., which is higher than the restoration threshold ($S_\mathrm{th}^{\mathrm{low}}=0.87$ p.u.), the system frequency drops again to 59.55 Hz to shed LG4. Once LG4 is shed, the BESS power drops below 0.87 p.u.. Thus, the system frequency is restored to 60 Hz, effectively ending the UFLS event. All shed loads will be restored after a recovery time of $\tau_\mathrm{rec}=15$ minutes.

The results show that the major drawback of the sectionalizer-based approach is that the disconnection of an entire LG may over-shed loads when meeting PRR. Moreover, the restoration of the shed LGs after 15-minute restoration timer may immediately trigger another low-power-reserve event. For instance, in Fig. \ref{fig:re_CB}, another UFLS event occurs around $t=2200$ s due to the restoration of LG4 and LG5.



\vspace{-10pt}
\subsection{Case 2: Appliances-based UFLS}
\begin{figure}[t]
	\centering
    \vspace{-10pt}
	\centerline{\includegraphics[width=0.43\textwidth]{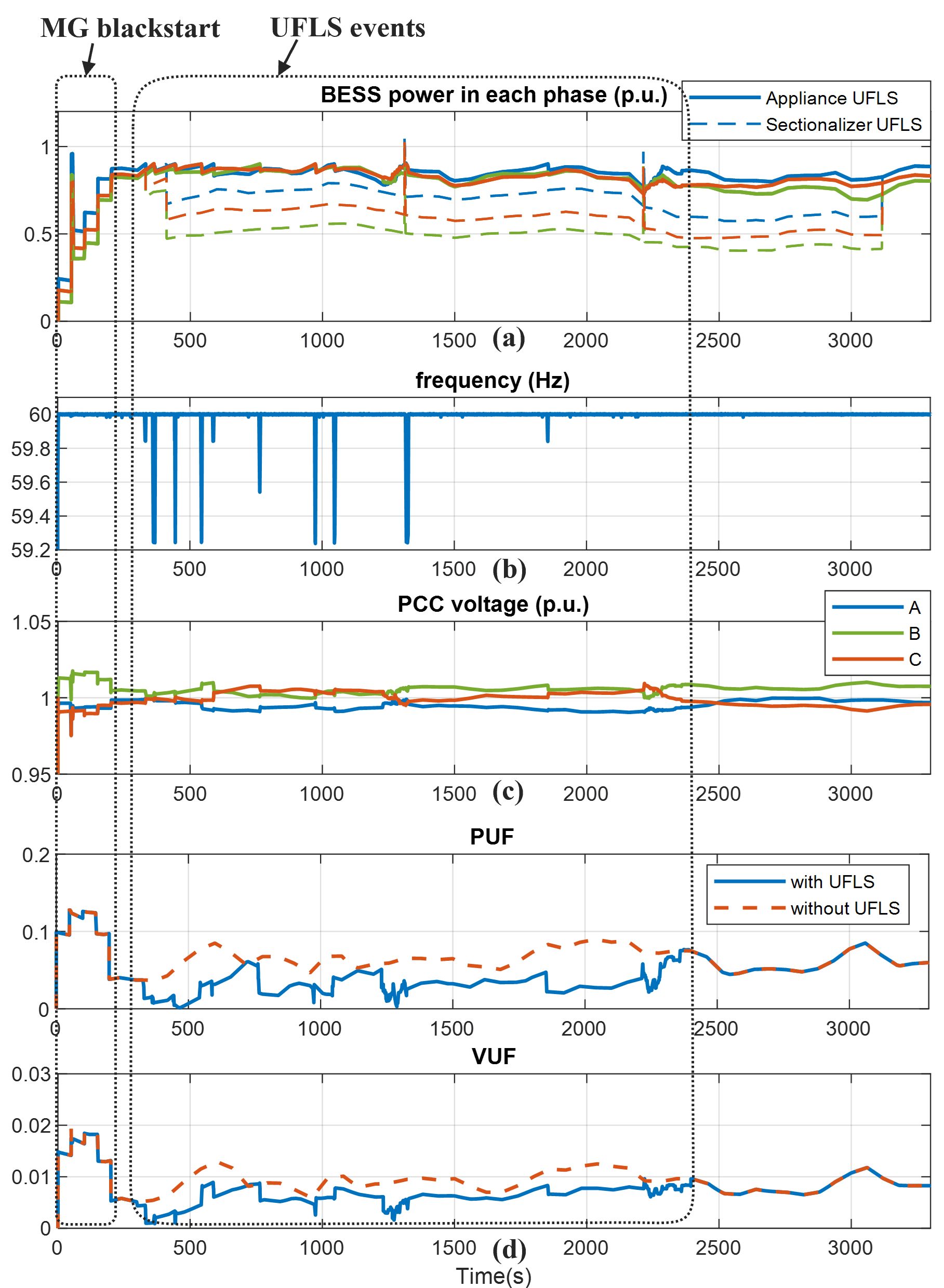}}
	\vspace{-10pt}
    \caption{Simulation results with appliance-level UFLS (a) BESS phase power; (b) System frequency; (c) PCC voltage; (d) Unbalance factor PUF and VUF.}
	\label{fig:re_app}
\end{figure}

Case 2 models the appliance-based UFLS scheme. The simulation results in comparison to the sectionalizer case are shown in Fig. \ref{fig:re_app}. From the results, we can highlight the following observations: 
\begin{itemize}[leftmargin=10px]
\setlength{\itemindent}{0.1mm}
    \item The BESS power output gradually decreases and recovers when fulfilling PRR, as opposed to the abrupt drop or immediate rebound observed in the sectionalizer-based scenario.
    \item Using appliance-based UFLS will lead to a higher number of UFLS events when compared to the sectionalizer case. This is because shedding controllable appliances across all LGs allows non-controllable loads to continue increasing. Note that preventing the disconnection of entire LGs can ensure equitable load shedding among LGs. 
    \item By examining the PUF and VUF plots, we can clearly see that per-phase UFLS has effectively alleviated the phase imbalance issue, in contrast to the sectionalizer case where the issue was exacerbated. This enhancement contributes to improved BESS control stability and MG operational stability, as excessive phase imbalance has the potential to trigger a complete shutdown of the entire MG.
\end{itemize}

\begin{figure}[t]
	\centering
	\centerline{\includegraphics[width=0.48\textwidth]{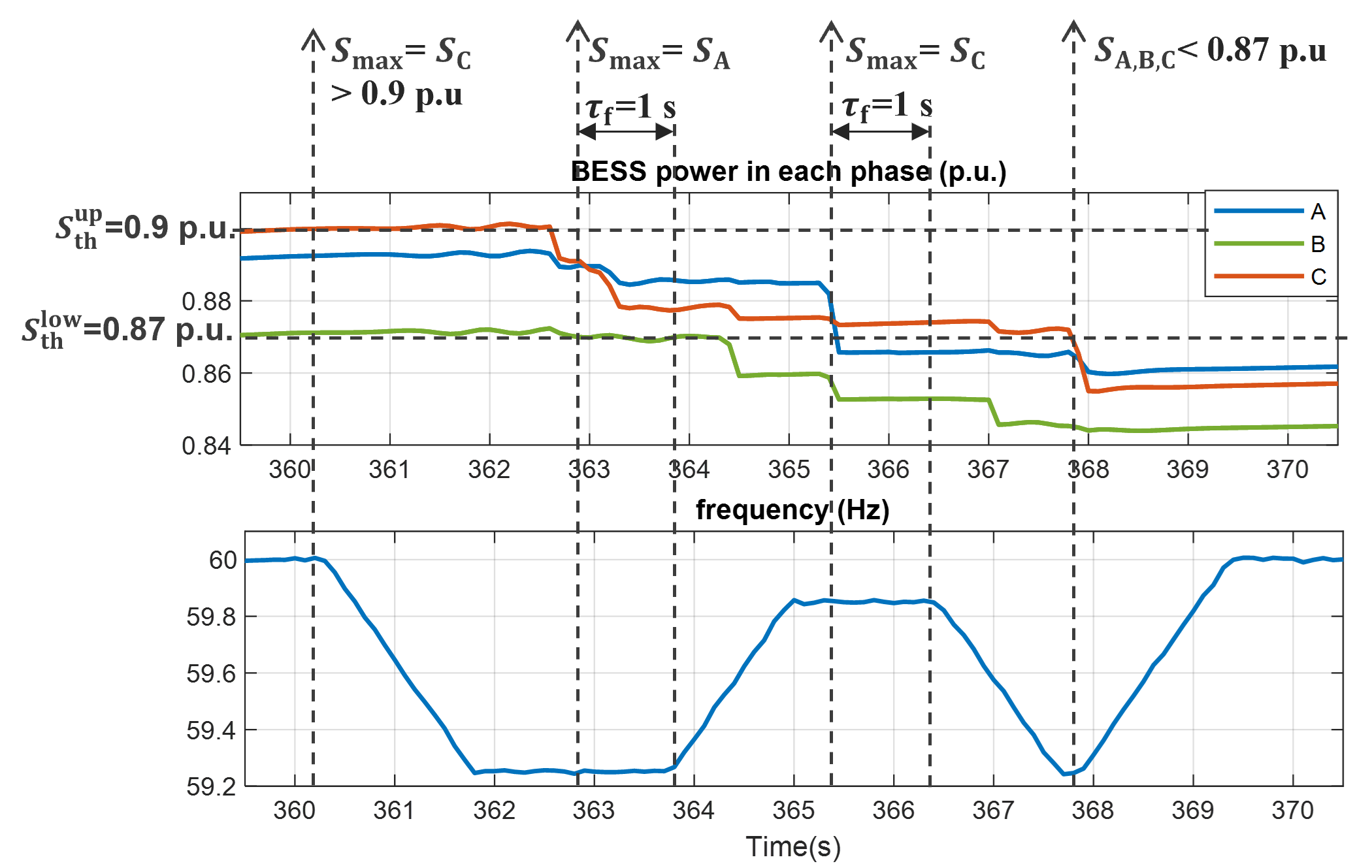}}
    \vspace{-10pt}
	\caption{Simulation results of an UFLS event with appliance-level UFLS.}
	\label{fig:re_app_in}
\end{figure} 

Fig. \ref{fig:re_app_in} presents a zoom-in plot of an UFLS event from 360 s to 370 s. The frequency setpoint is adapted according to the phase exhibiting the highest power output, ensuring that the load with the greatest power is shed, thereby contributing to effective three-phase imbalance mitigation.
At 360.2 s, UFLS is activated when the phase $c$ power exceeds the high power threshold, $S_\mathrm{C}>S_\mathrm{th}^\mathrm{up}=$ 0.9 p.u.. Since phase $c$ bears the highest load, the system frequency drops to the phase $c$ frequency setpoint stage, $f_\mathrm{C}=$ 59.25 Hz, with a 0.5 Hz/s ramp rate, shedding loads and hence decreasing the BESS phase $c$ output power.
When the phase with the highest power output switches from phase $c$ to phase $a$, the frequency will remains unchanged for a duration of $\tau_\mathrm{f}=1 s$ to validate the transition of the phase with maximum power output. 
Afterwards, the frequency is adjusted to phase $a$ setpoint, $f_\mathrm{A}=$ 59.85 Hz, to shed phase $a$ loads. This process persists until the power outputs across all three phases are lower than the low power threshold, $S_\mathrm{th}^\mathrm{low}=$ 0.87 p.u.. Upon reaching this state, the frequency ramps up to its nominal value of 60 Hz.

\begin{table}[t]
    \centering
        \vspace{-10pt}
    \caption{UFLS performance comparison: three-phase v.s. per-phase}\label{tab:re_compare}
    \vspace{-5pt}
    \begin{tabular}{C{3.6cm}|C{1.7cm}|C{1.7cm}}
        \hline
        \hline
         & 3-phase & Per phase \\
        \hline
        \hline
        PUF (compares to non-UFLS) & increase 0.1 & reduce 0.03 \\
        \hline
        VUF (compares to non-UFLS) & increase 1.1\% & reduce 0.2\% \\
        \hline
        Max. Deviation of $f$ (Hz) & 0.45  & 0.75 \\
        \hline
        Max. Deviation of $V_{\mathrm{pcc}}$ (p.u.) & 0.03 & 0.002 \\
        \hline
        Load Served (MWh) & 1.8  & 2.54  \\        
        \hline
        Number of UFLS device & 5 & 95 \\
        \hline
    \end{tabular}
\end{table}

\begin{figure}[b]
	\centering
	\centerline{\includegraphics[width=0.4\textwidth]{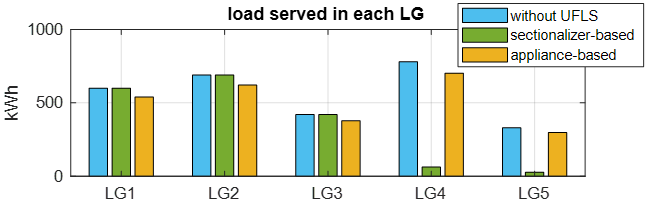}}
    \vspace{-11pt}
	\caption{UFLS performance comparison: load served in each load group.}
	\label{fig:re_LG}
\end{figure} 

To summarize, the simulation results show that both developed methods can effectively maintain power reserves for BESS and improve power system stability. However, The appliance-level UFLS method can reduce the shedding of excessive loads when compared to the sectionalizers-level UFLS method. Furthermore, the appliance-level UFLS scheme can better regulate the three-phase imbalance and avoid the potential imbalance issues caused by UFLS.

Table \ref{tab:re_compare} presents a summary of the observed metrics pertaining to the two proposed UFLS methodologies. The appliance-based method demonstrates superior performance by reducing both PUF and VUF, exhibiting lower frequency and voltage deviations, and enabling the shedding of fewer loads to meet the PRR. As indicated in Fig.~\ref{fig:re_LG}, a greater number of loads within LG4 and LG5 are being supplied. This illustrates that the per-phase based UFLS scheme effectively distributes the shedding of controllable loads across all LGs, preventing the repetitive shedding of LGs situated at the far end of the feeder (i.e., LG5 and LG4). This methodology ensures an equitable allocation of resources, promoting fairness in the provision of energy. However, this strategy does necessitate a larger number of controlled execution devices, thereby adding to the overall complexity of control.

Sectionalizers-level UFLS scheme can quickly shed whole loads in an area. However, it is non-selective, and may shed more loads than necessary. 
For the appliance-level UFLS scheme, the load shedding is selective and gradual. So it will cause smaller disturbances to the system. The disadvantage is that it requires controller chips to be installed in controllable loads (e.g. HVAC units). With the help of internet of things (IoT), the future grid can be composed of ``smart devices'' which include embedded UFLS functions.

\section{Conclusion}
This paper introduces a new UFLS scheme tailored for power reserve management in an islanded MG with a single GFM resource. The novel UFLS mechanism employs power reserve thresholds as event triggers and system frequency as control signals, achieving autonomous UFLS during regular operations. This reduces reliance on communication networks, crucial during extended main grid outages. By comparing per-phase UFLS to sectionalizer-based UFLS, we demonstrate the benefits in preventing excessive load shedding, addressing power imbalances, and ensuring equity when meeting PRR. The addition of triggering and recovery delays mitigates oscillations caused by transient events and synchronized load switching actions. Our future work will be focused on extending the proposed UFLS approach to broader grid services and MGs with multiple GFM resources.


%




\ifCLASSOPTIONcaptionsoff
  \newpage
\fi
\bibliographystyle{IEEEtran}
\bibliography{reference}


\end{document}